\documentstyle[twoside]{article}

\catcode`\@=11
\long\def\@makefntext#1{
\protect\noindent \hbox to 3.2pt {\hskip-.9pt  
$^{{\eightrm\@thefnmark}}$\hfil}#1\hfill}		

\def\@makefnmark{\hbox to 0pt{$^{\@thefnmark}$\hss}}	
	
\def\ps@myheadings{\let\@mkboth\@gobbletwo
\def\@oddhead{\hbox{}
\rightmark\hfil\eightrm\thepage}   
\def\@oddfoot{}\def\@evenhead{\eightrm\thepage\hfil
\leftmark\hbox{}}\def\@evenfoot{}
\def\sectionmark##1{}\def\subsectionmark##1{}}

\def\l{\lambda}
\def\beq{\begin{equation}}
\def\eeq{\end{equation}}

\oddsidemargin=\evensidemargin
\newcounter{sectionc}\newcounter{subsectionc}\newcounter{subsubsectionc}
\renewcommand{\section}[1] {\vspace{12pt}\addtocounter{sectionc}{1} 
\setcounter{subsectionc}{0}\setcounter{subsubsectionc}{0}\noindent 
	{\tenbf\thesectionc. #1}\par\vspace{5pt}}
\renewcommand{\subsection}[1] {\vspace{12pt}\addtocounter{subsectionc}{1} 
	\setcounter{subsubsectionc}{0}\noindent 
	{\bf\thesectionc.\thesubsectionc. {\kern1pt \bfit #1}}\par\vspace{5pt}}
\renewcommand{\subsubsection}[1] {\vspace{12pt}\addtocounter{subsubsectionc}{1}
	\noindent{\tenrm\thesectionc.\thesubsectionc.\thesubsubsectionc.
	{\kern1pt \tenit #1}}\par\vspace{5pt}}
\newcommand{\nonumsection}[1] {\vspace{12pt}\noindent{\tenbf #1}
	\par\vspace{5pt}}

\newcounter{appendixc}
\newcounter{subappendixc}[appendixc]
\newcounter{subsubappendixc}[subappendixc]
\renewcommand{\thesubappendixc}{\Alph{appendixc}.\arabic{subappendixc}}
\renewcommand{\thesubsubappendixc}
	{\Alph{appendixc}.\arabic{subappendixc}.\arabic{subsubappendixc}}

\renewcommand{\appendix}[1] {\vspace{12pt}
        \refstepcounter{appendixc}
        \setcounter{figure}{0}
        \setcounter{table}{0}
        \setcounter{lemma}{0}
        \setcounter{theorem}{0}
        \setcounter{corollary}{0}
        \setcounter{definition}{0}
        \setcounter{equation}{0}
        \renewcommand{\thefigure}{\Alph{appendixc}.\arabic{figure}}
        \renewcommand{\thetable}{\Alph{appendixc}.\arabic{table}}
        \renewcommand{\theappendixc}{\Alph{appendixc}}
        \renewcommand{\thelemma}{\Alph{appendixc}.\arabic{lemma}}
        \renewcommand{\thetheorem}{\Alph{appendixc}.\arabic{theorem}}
        \renewcommand{\thedefinition}{\Alph{appendixc}.\arabic{definition}}
        \renewcommand{\thecorollary}{\Alph{appendixc}.\arabic{corollary}}
        \renewcommand{\theequation}{\Alph{appendixc}.\arabic{equation}}
        \noindent{\tenbf Appendix \theappendixc #1}\par\vspace{5pt}}
\newcommand{\subappendix}[1] {\vspace{12pt}
        \refstepcounter{subappendixc}
        \noindent{\bf Appendix \thesubappendixc. {\kern1pt \bfit #1}}
	\par\vspace{5pt}}
\newcommand{\subsubappendix}[1] {\vspace{12pt}
        \refstepcounter{subsubappendixc}
        \noindent{\rm Appendix \thesubsubappendixc. {\kern1pt \tenit #1}}
	\par\vspace{5pt}}

\newcommand{\textlineskip}{\baselineskip=13pt}
\newcommand{\smalllineskip}{\baselineskip=10pt}

\def\eightcirc{
\begin{picture}(0,0)
\put(4.4,1.8){\circle{6.5}}
\end{picture}}
\def\eightcopyright{\eightcirc\kern2.7pt\hbox{\eightrm c}} 

\newcommand{\copyrightheading}[1]
	{\vspace*{-2.5cm}\smalllineskip{\flushleft
	{\footnotesize International Journal of Modern Physics A, #1}\\
	{\footnotesize $\eightcopyright$\, World Scientific Publishing
	 Company}\\
	 }}

\def\abstracts#1#2#3{{
	\centering{\begin{minipage}{4.5in}\baselineskip=10pt\footnotesize
	\parindent=0pt #1\par 
	\parindent=15pt #2\par
	\parindent=15pt #3
	\end{minipage}}\par}} 

\renewenvironment{thebibliography}[1]
	{\frenchspacing
	 \ninerm\baselineskip=11pt
	 \begin{list}{\arabic{enumi}.}
	{\usecounter{enumi}\setlength{\parsep}{0pt}
	 \setlength{\leftmargin 12.7pt}{\rightmargin 0pt} 
	 \setlength{\itemsep}{0pt} \settowidth
	{\labelwidth}{#1.}\sloppy}}{\end{list}}

\newcounter{itemlistc}
\newcounter{romanlistc}
\newcounter{alphlistc}
\newcounter{arabiclistc}

\newcommand{\fcaption}[1]{
        \refstepcounter{figure}
        \setbox\@tempboxa = \hbox{\footnotesize Fig.~\thefigure. #1}
        \ifdim \wd\@tempboxa > 5in
           {\begin{center}
        \parbox{5in}{\footnotesize\smalllineskip Fig.~\thefigure. #1}
            \end{center}}
        \else
             {\begin{center}
             {\footnotesize Fig.~\thefigure. #1}
              \end{center}}
        \fi}

\newcommand{\tcaption}[1]{
        \refstepcounter{table}
        \setbox\@tempboxa = \hbox{\footnotesize Table~\thetable. #1}
        \ifdim \wd\@tempboxa > 5in
           {\begin{center}
        \parbox{5in}{\footnotesize\smalllineskip Table~\thetable. #1}
            \end{center}}
        \else
             {\begin{center}
             {\footnotesize Table~\thetable. #1}
              \end{center}}
        \fi}

\def\@citex[#1]#2{\if@filesw\immediate\write\@auxout
	{\string\citation{#2}}\fi
\def\@citea{}\@cite{\@for\@citeb:=#2\do
	{\@citea\def\@citea{,}\@ifundefined
	{b@\@citeb}{{\bf ?}\@warning
	{Citation `\@citeb' on page \thepage \space undefined}}
	{\csname b@\@citeb\endcsname}}}{#1}}

\newif\if@cghi
\def\cite{\@cghitrue\@ifnextchar [{\@tempswatrue
	\@citex}{\@tempswafalse\@citex[]}}
\def\citelow{\@cghifalse\@ifnextchar [{\@tempswatrue
	\@citex}{\@tempswafalse\@citex[]}}
\def\@cite#1#2{{$\null^{#1}$\if@tempswa\typeout
	{IJCGA warning: optional citation argument 
	ignored: `#2'} \fi}}

\def\pmb#1{\setbox0=\hbox{#1}
	\kern-.025em\copy0\kern-\wd0
	\kern.05em\copy0\kern-\wd0
	\kern-.025em\raise.0433em\box0}

\def\fnt#1#2{\footnotetext{\kern-.3em
	{$^{\mbox{\scriptsize #1}}$}{#2}}}

\def\fpage#1{\begingroup
\voffset=.3in
\thispagestyle{empty}\begin{table}[b]\centerline{\footnotesize #1}
	\end{table}\endgroup}

\def\runninghead#1#2{\pagestyle{myheadings}
\markboth{{\protect\footnotesize\it{\quad #1}}\hfill}
{\hfill{\protect\footnotesize\it{#2\quad}}}}
\font\tenrm=cmr10
\font\tenit=cmti10 
\font\tenbf=cmbx10
\font\bfit=cmbxti10 at 10pt
\font\ninerm=cmr9

\font\eightrm=cmr8

\def\qed{\hbox{${\vcenter{\vbox{			
   \hrule height 0.4pt\hbox{\vrule width 0.4pt height 6pt
   \kern5pt\vrule width 0.4pt}\hrule height 0.4pt}}}$}}

\begin{document}

\runninghead{$\mu$ PROBLEM, SO(10) SUSY GUT AND HEAVY GLUINO LSP} 
{$\mu$ PROBLEM, SO(10) SUSY GUT AND HEAVY GLUINO LSP}

\normalsize\textlineskip
\thispagestyle{empty}
\setcounter{page}{1}

\copyrightheading{}			

\vspace*{0.88truein}

\fpage{1}
\centerline{\bf $\mu$ PROBLEM, SO(10) SUSY GUT AND HEAVY GLUINO LSP\footnote{This talk 
is based on the work done in collaboration with S. Raby. The
interested reader is referred to our paper \cite{MafiRaby3} for a more detailed discussion
of the model and its phenomenological implications.}
\footnote{Ohio state university preprint
OHSTPY-HEP-T-00-024.}}

\vspace*{0.37truein}
\centerline{\footnotesize ARASH MAFI}
\vspace*{0.015truein}
\centerline{\footnotesize\it Department of Physics, The Ohio
State University, 174 West 18th Ave.,}
\baselineskip=10pt
\centerline{\footnotesize\it Columbus, OH 43210, USA }
\vspace*{0.225truein}

\vspace*{0.21truein}
\abstracts{
We present a solution to the $\mu$ problem in an SO(10) supersymmetric
grand unified (SUSY GUT) model with gauge mediated (GMSB) and D-term supersymmetry 
breaking. A Peccei-Quinn ({\bf PQ}) symmetry is broken at the messenger scale and enables
the generation of the $\mu$ term. The invisible axion (Goldstone boson of {\bf PQ} 
symmetry breaking) is a cold dark matter candidate. At low energy, our model
leads to a phenomenologically acceptable version of the minimal supersymmetric 
standard model (MSSM) with novel particle phenomenology.
Either the gluino or the gravitino is the lightest supersymmetric particle (LSP).
The phenomenological constraints on the model result in a Higgs with 
mass $\sim 86 - 91$ GeV and $\tan\beta \sim 9 - 14$.}{}{}

\textlineskip			
\vspace*{12pt}			

\noindent

MSSM is a strongly motivated candidate for the physics beyond the
Standard Model (SM). There are two Higgs doublets ($H_u$ and $H_d$) 
in the MSSM. The vacuum condition obtained by minimizing the tree level Higgs
potential at the electroweak (EW) scale is 
\beq
\label{eq:musquared}   
\mu^2=-\frac{M_Z^2}{2}+
\frac{m^2_{H_d}- m^2_{H_u}tan^2\beta}{tan^2\beta-1},
\eeq
where $\tan{\beta}=\langle H_u\rangle/ \langle H_d\rangle$,
$m^2_{H_u}(m^2_{H_d})$ is the SSB mass of $H_u(H_d)$
and $B$ is the SSB Higgs bilinear coupling.
On the left hand side of Eq. \ref{eq:musquared}, the $\mu$ parameter,
multiplying a supersymmetric $\mu$ term in the Lagrangian, breaks no
SM symmetries; it could in principle be as large as the Planck or GUT scales.
On the right side, the $Z$ boson mass and the SSB Higgs masses are of order the
EW scale as required to solve the hierarchy problem in the SM. To avoid fine tuning in
Eq. \ref{eq:musquared}, the $\mu$ parameter should also be of order the
EW scale. The $\mu$ problem \cite{MafiRaby3} is stated as the difficulty to 
generate a $\mu$ parameter which is naturally of order the EW scale.

In this paper we use an extension of the GMSB model discussed
in Ref.\ \cite{RabyTobe3} to solve the $\mu$ problem. 
This model has GMSB with Higgs-messenger mixing in an SO(10) theory
and naturally leads to a stable gluino with mass in the
range $25 - 35$ GeV. Such gluino is still allowed by both the CDF 
and LEP data \cite{RabyMafi}$^,$\cite{BCG} and is a candidate for the 
ultra high energy cosmic rays (UHECRon). Either the gluino 
or the gravitino is the LSP in this
model. In the case of a gravitino LSP, gluino lives long enough (one month) to be
considered stable for both CDF-LEP and UHECRon analyses.   

The theory at the GUT scale is defined by the SO(10)
invariant superpotential $W\supset W_1+W_2+W_3$
and a non-renormalizable term in the Kahler
potential $K$ where

\begin{eqnarray}
\begin{array}{c}
W_1={\bf 16}_3 {\bf 10}_H {\bf 16}_3,\ \ \ \ \ \ \ \ \ \ \ \ \ \ \ \ \ \ \
W_2={\bf 10}_H A {\bf 10}_A+X {\bf 10}_A^2,\\ \\
W_3=\bar{\eta}_1 A\eta_1+
\bar{\eta}_2 A {\eta}_2+\bar{\bf \eta}_1{\bf \eta}_2,
\ \ \ \ \ \ \ \ \ \
K\supset \l_K\frac{X^\dagger}{M_P}{\bf 10}_H^2+h.c.
\end{array}
\end{eqnarray}
$({\bf 16}_3,\; \eta_1,\; \eta_2)$ are ${\bf 16}$'s,
$(\bar{\eta}_1,\; \bar{\eta}_2)$ are
$\bar{\bf 16}$'s,
$({\bf 10}_H,\; {\bf 10}_A)$ are ${\bf 10}$'s, 
$(X)$ is a singlet and $(A)$ is an adjoint under SO(10).

At the GUT scale, the theory is invariant
under a U(1) {\bf PQ} and an R symmetry.
The R symmetry is broken spontaneously at the GUT scale.
The {\bf PQ} symmetry, however, is not broken at the GUT scale 
and prevents a $\mu$ term in the superpotential.

$W_1$ contains the coupling of the third family matter multiplet (${\bf 16}_3$)
to the Higgs field  (${\bf 10}_H$) which includes  both the weak doublet and color triplet Higgs fields.

$W_2$ provides doublet-triplet splitting using the 
Dimopoulos-Wilczek mechanism in order to avoid
rapid proton decay \cite{DeMaRa}.
The adjoint field $A$ gets a vev $\langle A \rangle=(B-L)M_G$
where $B-L$ (baryon number minus lepton number) is non-vanishing
on color triplets and zero on weak doublets. The
singlet $X$ gets a vev $\langle X\rangle=M+\theta^2 F_X$.
This gives mass of order $M_G$ to the color triplet Higgs states and of order
$M$ to the weak doublets in ${\bf 10}_A$.   The Higgs doublets in ${\bf 10}_H$
remain massless. A SUSY breaking vev is also contained in $W_2$.

$W_2$ and $W_3$ provide the messengers for 
SUSY breaking.\footnote{Due to an accidental
cancellation, gluinos receive no mass at one loop from $W_2$.  Thus $W_3$ is
introduced with additional
messenger fields $(\eta_1,\bar{\eta}_1,\eta_2,\bar{\eta}_2)$ contributing
to the masses of gauginos and scalars at the scale $M_G$.}
The auxiliary field ${\bf 10}_A$ and the fields $\bar \eta_1, \eta_2$ feel SUSY
breaking at tree level due to the vev $F_X$.   They are thus
the messengers for GMSB.
We take the messenger scale $M\sim 10^{12}$ GeV with the effective SUSY breaking
scale in the observable sector given by $\Lambda=F_X/M\sim 10^5\ GeV$.

When $X$ gets a vev, both SUSY and the {\bf PQ} symmetry
are broken.   The $\mu$ term is generated at the scale $M$, $\mu=\l_K\frac{F_X}{M_P}$,
while $B$ remains zero at tree level \footnote{It is necessary to generate
$B$ at higher loop orders than $\mu$ to avoid a $B$ hierarchy problem.}.

The {\bf PQ} symmetry solves the strong CP problem and produces an axion;
the Goldstone boson of the broken {\bf PQ} symmetry. Such axion is a candidate
for the cold dark matter.

The boundary conditions at the messenger scale are determined
by two sources of SUSY breaking, gauge 
mediation and D-term \cite{MafiRaby3,RabyTobe3}.
The messengers give mass to the gauginos and Higgs at one loop and 
to squarks and sleptons at two loops.  Since the color triplet messengers 
have mass of order the GUT scale, the gluino mass is suppressed compared to the other
gauginos. For phenomenological reasons we assume that SUSY
is also broken by the D-term of an anomalous $U(1)_X$ gauge
symmetry \cite{MafiRaby3,RabyTobe3}.  Moreover, the GMSB and D-term 
contributions are necessarily 
comparable \cite{RabyTobe4}. The D-term contribution to scalar masses is given
by $\delta_D\tilde{m}^2_a= d \; Q^X_a \; M_2^2$,
where $Q^X_a$ is the $U(1)_X$ charge of the field $a$ and
$d$ is an arbitrary parameter of order $1$ which measures  
the strength of D-term versus gauge-mediated SUSY breaking.
The value of $Q^X_a$ for $a = {\bf 16},\;  {\bf 10},\; {\bf 1}$
of SO(10) is given by $1,\; -2, \;4$ \cite{MafiRaby3,RabyTobe3}.

Imposing gauge coupling unification at the 
GUT scale, we renormalize the effective Lagrangian parameters to the
EWSB scale and determine the free parameters of our model by fitting the low energy data 
which we take to include $m_t,\; m_b,\; m_\tau,\; \alpha_{em},\; \alpha_{s}$ and 
$\sin^2{\theta_W}$.
 
We now consider the LEP constraints on parameters in
our model.  The most important constraints come from the latest
Higgs search results at LEP \cite{Cernlatest}.
In our model the off-diagonal elements in the stop mass-squared matrix
are very small, thus the severe LEP limits for the neutral Higgs in 
the no stop-quark mixing scenario are most 
applicable \cite{Cernlatest}. In our model only $d\sim\ 0.40-0.45$ survives the 
LEP constraint for $\Lambda=10^5$ GeV. With these values of 
the parameters, the mass of the lightest neutral Higgs
resides in the narrow range $\sim 86 - 91$ GeV with $\tan\beta \sim 9 - 14$.
For $d\sim\ 0.40-0.45$ and $\Lambda=10^5$ GeV we find 
the lightest stop and neutralino with mass in
the range 100 - 122 GeV and  50 - 72 GeV, respectively\footnote{Note, a 
recent LEP bound on a heavy gluino LSP using stop production and 
decay \cite{katsenevas} does not constrain the model
since the stop mass in our case is larger than the values probed in this search.}.

Our model survives the OPAL limit on $e^+\ e^-\to\ $hadrons coming 
from chargino and neutralino
pair production. We have also analyzed the one loop branching 
ratio for $b\to s\gamma$. We suggest that in the desired region of parameter
space, our model is not excluded by CLEO data.

To summarize, we have presented a solution to the $\mu$ and strong
CP problems in the presence of a heavy gluino LSP.
Either the gluino or the gravitino is the LSP. A light gluino
reduces the fine tuning necessary for EWSB and is a candidate
for the UHECRon. It also leads to a model with Higgs mass 
of order 90 GeV and a stop mass less than the top satisfying 
some of the dynamical constraints necessary for electroweak 
baryogenesis in supersymmetric theories. The axion; the Goldstone
boson of the {\bf PQ} symmetry is also a candidate for the cold dark matter.

\nonumsection{Acknowledgements}
\noindent
I would like to thank my collaborator professor S. Raby and organizers
of the DPF2000. This work was supported in part by the DOE/ER/01545-796..

\nonumsection{References}

\end{document}